\documentclass[letterpaper,english,twocolumn]{revtex4-2}
\usepackage[T1]{fontenc}
\usepackage[latin9]{inputenc}
\usepackage{amsmath}
\usepackage{amssymb}
\usepackage{graphicx}

\makeatletter

\providecommand{\tabularnewline}{\\}

\makeatother

\usepackage{babel}
\begin{document}
\title{Operator overlaps in harmonic oscillator bases with different oscillator
lengths}
\author{L.M. Robledo}
\email{luis.robledo@uam.es}

\affiliation{Departamento de Física Teórica and CIAF, Universidad Autónoma de Madrid,
E-28049 Madrid, Spain}
\affiliation{Center for Computational Simulation, Universidad Politécnica de Madrid,
Campus de Montegancedo, Boadilla del Monte, E-28660-Madrid, Spain}
\date{\today}
\begin{abstract}
We apply a formalism recently developed to carry out Generator Coordinate
Method calculations using a set of Hartree- Fock- Bogoliubov wave
functions, where each of the members of the set can be expanded in
an arbitrary basis. In this paper it is assumed that the HFB wave
functions are expanded in Harmonic Oscillator (HO) bases with different
oscillator lengths. General expressions to compute the required matrix
elements of arbitrary operators are given. The application of the
present formalism to the case of fission is illustrated with an example.
\end{abstract}
\maketitle

\section{\label{sec:Intro}Introduction}

Based on previous results published in Ref \citep{PhysRevC.50.2874}
we have developed in a recent publication \citep{Rob22} (denoted
by I in the following) a formulation of the generalized Wick's theorem
to compute overlaps between Hartree-Fock-Bogoliubov (HFB) wave functions
expressed in different bases not connected by unitary transformations.
In this paper we apply the formalism to the common situation where
one deals with two finite harmonic oscillator (HO) bases with different
oscillator lengths. The present results could be of interest to carry
out generator coordinate method (GCM) calculations and/or symmetry
restoration \citep{ring2000,bender2003,Robledo2019,Sheikh2021} in
fission - see \citep{Schunck2016,Marevic2020} for a recent review
and for a recent application. In fission one has to consider a set
of HFB wave functions usually labeled in terms of the axial quadrupole
moment $|\phi(q_{20})\rangle$ to study fission dynamics. The set
of HFB states $|\phi(q_{20})\rangle$ span a large and rich set of
nuclear shapes (from deformed ground states to configurations with
a thin neck ending up finally in scission configurations) and therefore
the oscillator lengths of the HO bases used have to adapt to the corresponding
shape as to reduce/optimize basis size. Typically, the oscillator
lengths for each quadrupole moment are determined by minimizing the
HFB energy and the range of values obtained in a typical fission path
can be rather large. As a consequence of the different oscillator
lengths used for each $q_{20}$value the bases used to express the
corresponding HFB wave functions are not connected by unitary transformations
(as a consequence of being finite dimensional). Therefore calculation
of norm and Hamiltonian overlaps between the different configurations
cannot be carried out with the traditional formulas \citep{Onishi1966367,Balian.69}
as they assume bases connected by unitary transformation and therefore
a generalization of Wick's theorem for overlaps between general HFB
states is required. At this point it has to be mentioned that the
evaluation of overlaps between Slater determinants \citep{PhysRev.97.1490}
is properly handled even if the bases are not connected by unitary
transformation. The most straightforward solution to this problem
would be to use a common basis (with the same oscillator lengths)
for all the HFB states in the fission path but this solution demands
huge harmonic oscillator bases and humongous computational resources.
At this point the reader might also wonder if it would not be better
to use an unique, large enough basis, for all the relevant HFB states
as it is usually done in calculations in the mesh. This alternative
is however impractical for all kind of interactions except for those
of zero range with trivial local exchange terms. In order to find
a practical solution to the above problems a generalization of the
method used in \citep{Onishi1966367,Balian.69} was formulated in
Ref \citep{PhysRevC.50.2874}. The solution to all these problems
relies on the formal extension of the original bases as to make them
complete and therefore unitarily connected. The basis states to be
added come with zero occupancy. This approach has been pursued in
Refs \citep{BONCHE1990466,Valor2000145} for unitary transformations
and in Ref \citep{PhysRevC.50.2874} for general canonical ones. Years
later, the pfaffian formula for the proper calculation of the norm
overlaps including its sign \citep{PhysRevC.79.021302,PhysRevLett.108.042505}
was subsequently generalized as to consider different bases \citep{PhysRevC.84.014307}.
The problem with the formulation of Ref \citep{PhysRevC.50.2874}
was that additional considerations were required to come to the final
formulas. Recently, in Ref \citep{Rob22} we have reformulated the
work of Ref \citep{PhysRevC.50.2874} as to simplify the expressions
for the contractions entering the Hamiltonian and other operators
overlaps. The main advantage of the new formulation is that it becomes
evident that the operator's overlap can be obtained in terms of what
we will call intrinsic quantities (i.e. quantities that can be solely
computed within the given finite bases). In this paper we apply the
formalism to the above mentioned situation of HO bases with different
oscillator lengths. The required overlap matrix between the two HO
bases with different lengths is explicitly built as well as its Lower
Upper (LU) decomposition that play a central role in the final expressions
of Ref \citep{Rob22}. The main application of the present formulation
is to carry out GCM calculations or restore spatial symmetries in
a fission framework. The new formulas could also be used to provide
a more robust and precise formulation of the Time Dependent GCM. This
will definitely help to improve our understanding of fission fragment
properties and fission dynamics.

\section{The generalized Wick theorem for arbitrary basis}

Recently \citep{Rob22}, a convenient formalism to evaluate the overlap
of general multi-body operators between arbitrary HFB wave functions
\begin{equation}
\frac{\langle\phi_{0}|\hat{O}|\phi_{1}\rangle}{\langle\phi_{0}|\phi_{1}\rangle}\label{eq:over1}
\end{equation}
was laid down. In it, each of the HFB states entering the overlap
are expanded in different bases not connected by unitary transformations
(i.e. not expanding the same subspace of the whole Hilbert space).
First, we establish the notation and then present the main results.
The bases and associated creation operators are denoted by $\mathcal{B}_{0}=\{c_{0,k}^{\dagger},k=1,\ldots,N_{0}\}$
in the case of $|\phi_{0}\rangle$ and $\mathcal{B}_{1}=\{c_{1,k}^{\dagger},k=1,\ldots,N_{1}\}$
in the case of $|\phi_{1}\rangle$. Both bases satisfy canonical fermion
anti-commutation relations (CAR) , i.e $\{c_{i,k},c_{i,k'}\}=\delta_{kk'}$
and are connected by an overlap matrix $\{c_{0,k}^{\dagger},c_{1,l}\}=_{0}\langle k|l\rangle_{1}=\mathcal{R}_{kl}$.
For simplicity, we assume in the following $N_{0}=N_{1}=N$, but note
that the most general case can be easily accommodated in the formalism.
We will also introduce the complement of the two bases $\bar{\mathcal{B}}_{0}=\{c_{0,k}^{\dagger},k=N+1,\ldots,\infty\}$
and $\bar{\mathcal{B}}_{1}=\{c_{1,k}^{\dagger},k=N+1,\ldots,\infty\}$
such that $\mathcal{B}_{0}\cup\bar{\mathcal{B}}_{0}=\{c_{0,k}^{\dagger}\}^{\infty}$
and $\mathcal{B}_{1}\cup\bar{\mathcal{B}}_{1}=\{c_{1,k}^{\dagger}\}^{\infty}$
expand the whole separable Hilbert space and therefore represent bases
connected by a unitary (infinite dimensional) transformation matrix
$R$ (not to be confused with $\mathcal{R}$). Let us also introduce
the quasiparticle annihilation operators $\alpha_{i\mu}$ ($i=0,1)$,
which annihilate $|\phi_{i}\rangle$. They are written in terms of
the complete bases $\{c_{i,k}^{\dagger}\}^{\infty}$ through the standard
definition
\[
\alpha_{i\mu}=\sum_{k}\left(U_{i}^{*}\right)_{k\mu}c_{i,k}+\left(V_{i}^{*}\right)_{k\mu}c_{i,k}^{\dagger}.
\]
The Bogoliubov amplitudes $U_{i}$ and $V_{i}$ have a block structure
\[
V_{i}=\left(\begin{array}{cc}
\bar{V}_{i} & 0\\
0 & 0
\end{array}\right),\;\;U_{i}=\left(\begin{array}{cc}
\bar{U}_{i} & 0\\
0 & d_{i}
\end{array}\right),
\]
where $\bar{V}_{i}$ and $\bar{U}_{i}$ are the $N\times N$ matrices
characterizing $|\phi_{i}\rangle$. In this way, the set of N quasiparticle
operators $\alpha_{i\mu}$ with $\mu=1,\ldots,N$, correspond to the
quasiparticle operators expanded in the truncated bases $\mathcal{B}_{i}$.
The $d_{i}$ are arbitrary unitary matrices that play no role in the
final expressions. It is also convenient to express the unitary matrix
$R$ connecting $\mathcal{B}_{0}\cup\bar{\mathcal{B}}_{0}$ with $\mathcal{B}_{1}\cup\bar{\mathcal{B}}_{1}$
as a block matrix
\[
R=\left(\begin{array}{cc}
\mathcal{R} & \mathcal{S}\\
\mathcal{T} & \mathcal{U}
\end{array}\right).
\]
The matrix $R$ is just the representation of the unitary operator
$\hat{\mathcal{T}}_{01}$ connecting the two complete  bases $\hat{\mathcal{T}}_{01}c_{0,k}^{\dagger}\hat{\mathcal{T}}_{01}^{\dagger}=c_{1,k}^{\dagger}$
whereas $\mathcal{R}$ is the restriction of this operator to the
bases $\mathcal{B}_{0}$ and $\mathcal{B}_{1}$. In the present case,
where we are dealing with HO bases differing in their oscillator lengths,
$\hat{\mathcal{T}}_{01}$ is just the dilatation operator. As discussed
in Appendix \ref{sec:The-dilatation-operator}, it is the exponential
of an one body operator. As shown in Ref \citep{Rob22} the calculation
of the overlap of Eq. (\ref{eq:over1}) simplifies enormously if the
operator $\hat{O}$ is written in second quantization form in terms
of both bases $\{c_{0,k}^{\dagger}\}^{\infty}$ and $\{c_{1,k}^{\dagger}\}^{\infty}$
. One-body operators are expressed in the form 
\begin{equation}
\hat{O}=\sum_{kl}O_{kl}^{01}c_{0,k}^{\dagger}c_{1,l}\label{eq:onebody}
\end{equation}
with 
\[
O_{kl}^{01}=_{0}\langle k|\hat{O}|l\rangle_{1}.
\]
In the same way a two body operator will be expressed as
\begin{equation}
\hat{O}=\frac{1}{4}\sum_{k_{1}k_{2}l_{1}l_{2}}\tilde{\upsilon}_{k_{1}k_{2}l_{1}l_{2}}^{01}c_{0k_{1}}^{\dagger}c_{0,k_{2}}^{\dagger}c_{1,l_{2}}c_{1,l_{1}}\label{eq:Twobody}
\end{equation}
where the antisymmetrized two body matrix element is given by $\tilde{\upsilon}_{k_{1}k_{2}l_{1}l_{2}}^{01}=\upsilon_{k_{1}k_{2}l_{1}l_{2}}^{01}-\upsilon_{k_{1}k_{2}l_{2}l_{1}}^{01}$
and $\upsilon_{k_{1}k_{2}l_{2}l_{1}}^{01}=_{0}\langle k_{1}k_{2}|\hat{\upsilon}|l_{1}l_{2}\rangle_{1}$
are the interaction's matrix elements. The extension to higher order
operators is straightforward. The sums in Eqs (\ref{eq:onebody})
and (\ref{eq:Twobody}) extend over the complete bases $\{c_{0,k}^{\dagger}\}^{\infty}$
or $\{c_{1,k}^{\dagger}\}^{\infty}$ to faithfully represent the operators.
As shown in I, the overlaps of those operators can be obtained by
using the standard rules of Wick's theorem but using the elementary
contractions 
\begin{align}
\rho_{lk}^{01} & =\frac{\langle\phi_{0}|c_{0,k}^{\dagger}c_{1,l}|\phi_{1}\rangle}{\langle\phi_{0}|\phi_{1}\rangle}\label{eq:rho10F}\\
 & =\begin{cases}
\left[\bar{V_{1}^{*}}A^{-1}\bar{V}_{0}^{T}\right]_{lk} & l\subset\mathcal{B}_{0},k\subset\mathcal{B}_{1}\\
0 & \textrm{otherwise}
\end{cases}\nonumber \\
\bar{\kappa}_{k_{1}k_{2}}^{01} & =\frac{\langle\phi_{0}|c_{0,k_{1}}^{\dagger}c_{0,k_{2}}^{\dagger}|\phi_{1}\rangle}{\langle\phi_{0}|\phi_{1}\rangle}\label{eq:kappa01F}\\
 & =\begin{cases}
-\left[\left(\mathcal{R}^{T}\right)^{-1}\bar{U}_{1}^{*}A^{-1}\bar{V}_{0}^{T}\right]_{k_{1}k_{2}} & k_{1}\subset\mathcal{B}_{0},k_{2}\subset\mathcal{B}_{0}\\
0 & \textrm{otherwise}
\end{cases}\nonumber \\
\kappa_{l_{1}l_{2}}^{10} & =\frac{\langle\phi_{0}|c_{1,l_{1}}c_{1,l_{2}}|\phi_{1}\rangle}{\langle\phi_{0}|\phi_{1}\rangle}\label{eq:kappa10F}\\
 & =\begin{cases}
\left[\bar{V}_{1}^{*}A^{-1}\bar{U}_{0}^{T}\left(\mathcal{R}^{T}\right)^{-1}\right]_{l_{1}l_{2}} & l_{1}\subset\mathcal{B}_{1},l_{2}\subset\mathcal{B}_{1}\\
0 & \textrm{otherwise}
\end{cases}\nonumber 
\end{align}
and therefore all the indices in the sums are restricted to the subspace
spanned by$\mathcal{B}_{0}$ or $\mathcal{B}_{1}$. Although not obvious
from the expressions in the right hand side of Eqs (\ref{eq:kappa01F})
and (\ref{eq:kappa10F}) the quantities $\bar{\kappa}_{k_{1}k_{2}}^{01}$
and $\kappa_{l_{2}l_{1}}^{10}$ are skew-symmetric matrices. In the
above expressions, the matrix $A$ is given by
\begin{equation}
A=\bar{U}_{0}^{T}\left(\mathcal{R}^{T}\right)^{-1}\bar{U}_{1}^{*}+\bar{V}_{0}^{T}\mathcal{R}\bar{V}_{1}^{*}.\label{eq:Amatrix}
\end{equation}
 Using the above contractions the overlap of an one-body operator
is given by 
\begin{equation}
\frac{\langle\phi_{0}|\hat{O}|\phi_{1}\rangle}{\langle\phi_{0}|\phi_{1}\rangle}=\sum_{k,l=1}^{N}O_{kl}^{01}\rho_{lk}^{01}=\textrm{Tr}[O^{01}\rho^{01}]\label{eq:OneBodyOverlap}
\end{equation}
whereas for a two body one we have
\begin{align}
\frac{\langle\phi_{0}|\hat{O}|\phi_{1}\rangle}{\langle\phi_{0}|\phi_{1}\rangle} & =\frac{1}{4}\sum_{k_{1},k_{2},l_{1},l_{2}=1}^{N}\tilde{\upsilon}_{k_{1}k_{2}l_{1}l_{2}}^{01}[\rho_{l_{1}k_{1}}^{01}\rho_{l_{2}k_{2}}^{01}\label{eq:Twobodyoverlap}\\
 & -\rho_{l_{1}k_{2}}^{01}\rho_{l_{2}k_{1}}^{01}+\bar{\kappa}_{k_{1}k_{2}}^{01}\kappa_{l_{2}l_{1}}^{10}]
\end{align}
Please remember that contrary to the standard method, the matrices
$O_{kl}^{01}$, $\rho_{lk}^{01}$ and $\tilde{\upsilon}_{k_{1}k_{2}l_{1}l_{2}}^{01}$
are not hermitian but the matrices $\bar{\kappa}_{k_{1}k_{2}}^{01}$
and $\kappa_{l_{1}l_{2}}^{10}$are still skew-symmetric. The overlap
of the two HFB wave functions is 
\begin{equation}
\langle\phi_{0}|\phi_{1}\rangle=\sqrt{\det A\det\mathcal{R}}.\label{eq:Overlap}
\end{equation}
In Ref \citep{Rob22} a subsequent Lower-Upper (LU) decomposition
of $\mathcal{R}$ was introduced
\begin{equation}
\mathcal{R}=L_{0}^{*}L_{1}^{T}\label{eq:LUR}
\end{equation}
where $L_{0}$ and $L_{1}$ are both lower triangular matrices. The
decomposition introduces implicitly a bi-orthogonal basis $|k)_{1}=\sum\left(L_{1}^{T}\right)_{jk}^{-1}|j\rangle_{1}$
and $_{0}(l|=\sum_{0}\langle i|\left(L_{0}^{*}\right)_{li}^{-1}$
such that $_{0}(l|k)_{1}=\delta_{lk}.$ The LU decomposition of the
overlap matrix suggests the definitions
\begin{align}
\tilde{U}_{0}=\left(L_{0}^{*}\right)^{-1}\bar{U}_{0}L_{0}^{+} & \qquad\tilde{V}_{0}=L_{0}^{+}\bar{V}_{0}L_{0}^{+}\label{eq:U0tilde}\\
\tilde{U}_{1}=\left(L_{1}^{*}\right)^{-1}\bar{U}_{1}L_{1}^{+} & \qquad\tilde{V}_{1}=L_{1}^{+}\bar{V}_{1}L_{1}^{+}\label{eq:U1tilde}
\end{align}
that turn out to be very useful to define handy quantities not depending
explicitly on $\mathcal{R}$ like, for instance,
\begin{equation}
\tilde{A}=\tilde{U}_{0}^{T}\tilde{U}_{1}^{*}+\tilde{V}_{0}^{T}\tilde{V}_{1}^{*}=L_{0}^{*}AL_{1}^{T}.\label{eq:Atilde}
\end{equation}
The overlap is now written as
\begin{equation}
\langle\phi_{0}|\phi_{1}\rangle=\sqrt{\det\tilde{A}}\label{eq:Overtilde}
\end{equation}
It is also convenient to introduce the contractions
\begin{align}
\tilde{\rho}_{lk}^{01} & =\left[\tilde{V}_{1}^{*}\tilde{A}^{-1}\tilde{V}_{0}^{T}\right]_{lk}=L_{1}^{T}\rho^{01}L_{0}^{*}\label{eq:rho10F-1}\\
\tilde{\bar{\kappa}}_{k_{1}k_{2}}^{01} & =-\left[\tilde{U}_{1}^{*}\tilde{A}^{-1}\tilde{V}_{0}^{T}\right]_{k_{1}k_{2}}=L_{0}^{+}\bar{\kappa}{}^{01}L_{0}^{*}\label{eq:kappa01F-1}\\
\tilde{\kappa}_{l_{1}l_{2}}^{10} & =\left[\tilde{V}_{1}^{*}\tilde{A}^{-1}\tilde{U}_{0}^{T}\right]_{l_{1}l_{2}}=L_{1}^{T}\kappa{}^{01}L_{1}\label{eq:kappa10F-1}
\end{align}
Using them and the matrix elements $\tilde{O}=\left(L_{0}^{*}\right)^{-1}O^{01}\left(L_{\text{1}}^{T}\right)^{-1}$
one gets $\mathrm{Tr}(\tilde{O}\tilde{\rho}^{01})$ for the overlap
of an one-body operator. Please note that $\tilde{O}_{lk}$ are the
matrix elements of the operator $\hat{O}$ in the bi-orthogonal basis
$_{0}(l|$ and $|k)_{1}$, i.e. $\tilde{O}_{lk}=_{0}(l|\hat{O}|k)_{1}=\sum_{ij}\left(L_{0}^{*}\right)_{li}^{-1}{}_{0}\langle i|\hat{O}|j\rangle_{1}\left(L_{1}^{T}\right)_{jk}^{-1}$.
Similar considerations apply to the overlap of two-body operators.
Introducing the two-body matrix element in the bi-orthogonal basis
$\upsilon_{ijkl}^{B}=\mbox{}_{0}(ij|\hat{\upsilon}|kl)_{1}$, related
to $\upsilon_{ijkl}^{01}$ by 
\[
\upsilon^{B}=\left(L_{0}^{*}\right)^{-1}\left(L_{0}^{*}\right)^{-1}\upsilon^{01}\left(L_{1}^{T}\right)^{-1}\left(L_{1}^{T}\right)^{-1}
\]
we can define HF potential $\tilde{\Gamma}_{ik}^{01}=\frac{1}{2}\sum\tilde{\upsilon}_{ijkl}^{B}\tilde{\rho}_{lj}^{01}$
and pairing field $\tilde{\Delta}_{ij}^{01}=\frac{1}{2}\sum\tilde{\upsilon}_{ijkl}^{B}\tilde{\kappa}_{kl}^{01}$
to write 
\begin{equation}
\frac{\langle\phi_{0}|\hat{\upsilon}|\phi_{1}\rangle}{\langle\phi_{0}|\phi_{1}\rangle}=\frac{1}{2}\mathrm{Tr}[\tilde{\Gamma}^{01}\tilde{\rho}^{01}]-\frac{1}{2}\mathrm{Tr}[\tilde{\Delta}^{01}\tilde{\bar{\kappa}}^{01}]\label{eq:mv_v}
\end{equation}
which is again the standard expression but defined in terms of Eqs
(\ref{eq:rho10F-1}), (\ref{eq:kappa01F-1}) and (\ref{eq:kappa10F-1})
and the definitions above. The advantage of the definitions in Eqs
(\ref{eq:Atilde}), (\ref{eq:rho10F-1}), (\ref{eq:kappa01F-1}) and
(\ref{eq:kappa10F-1}) is that they have exactly the same expression
as the formulas available in the literature for complete bases but
expressed in terms of the ``tilde'' $U$ and $V$ matrices of Eqs
(\ref{eq:U0tilde}) and (\ref{eq:U1tilde}). There is an additional
advantage in the fact that $\tilde{A}$ is a ``more balanced'' matrix
being less affected by the near singular character of the overlap
matrix $\mathcal{R}.$ 

\section{Application to HO bases with different oscillator lengths }

To apply the above formalism to the case of HO bases with different
oscillator lengths we just need to compute the overlap matrix $\mathcal{R}_{kl}={}_{0}\langle k|l\rangle_{1}$
and the overlap matrix elements $O_{kl}^{01}={}_{0}\langle k|\hat{O}|l\rangle_{1}$
and $_{0}\langle k_{1}k_{2}|\hat{\upsilon}|l_{1}l_{2}\rangle_{1}$
for the harmonic oscillator bases with different oscillator lengths.
To simplify the discussion, I will restrict to the case of a HO basis
tensor product of 1D states $\varphi_{n}(\vec{r})=\prod_{i=1}^{3}\varphi_{n_{i}}(x_{i},b_{i})$
with $\varphi_{n_{i}}(x_{i},b_{i})=e^{-1/2x_{i}^{2}/b_{i}^{2}}\bar{\varphi}_{n_{i}}(x_{i},b_{i})$
the product of a Gaussian factor times a polynomial 
\[
\bar{\varphi}_{n_{i}}(x_{i},b_{i})=1/\sqrt{\sqrt{\pi}2^{n_{i}}n_{i}!b_{i}}H_{n_{i}}\left(x_{i}/b_{i}\right)
\]
proportional to the Hermite polynomial $H_{n}$ of degree $n$. As
any polynomial of degree $n$ can be written as a linear combination
of $n+1$ polynomials of degree $n$ or less, we can express $\bar{\varphi}_{n}(x,b_{0})$
in terms of $\bar{\varphi}_{n}(x,b_{1})$ by means of a finite dimensional
lower triangular transformation matrix $L_{nm}(q_{01})$ that depends
on the ratio $q_{01}=b_{1}/b_{0}$ (see Appendix \ref{sec:AppendixB})
\[
\bar{\varphi}_{n}(x,b_{0})=\sum_{m=0}^{n}L_{nm}(q_{01})\bar{\varphi}_{m}(x,b_{1})
\]
Due to the lower triangular structure of the matrix $L$ both its
inverse and determinant can be obtained analytically (see Appendix
\ref{sec:AppendixB}). It is now straightforward to compute the 1D
overlaps
\begin{align*}
\mathcal{R}_{nm} & =\int dx\:\varphi_{n}^{*}(x,b_{0})\varphi_{m}(x,b_{1})\\
 & =\int dx\:e^{-\frac{x^{2}}{B^{2}}}\bar{\varphi}_{n}^{*}(x,b_{0})\bar{\varphi}_{m}(x,b_{1})\\
 & =\sum_{r}L_{nr}^{*}(q_{0})L_{mr}(q_{1})
\end{align*}
where the new oscillator length $B$ is given by 
\begin{equation}
1/B^{2}=\frac{1}{2}\left(1/b_{0}^{2}+1/b_{1}^{2}\right)\label{eq:Bosc}
\end{equation}
and one has introduced the parameters $q_{0}=B/b_{0}$ and $q_{1}=B/b_{1}$.
The matrices $L(q_{i})$ are the ones transforming $\bar{\varphi}_{m}(x,b_{i})$
into $\bar{\varphi}_{m}(x,B)$. The full overlap matrix is then given
by 
\begin{equation}
\mathcal{R}_{nm}={}_{0}\langle n|m\rangle_{1}=\left(L^{*}(\mathbf{q}_{0})L^{T}(\mathbf{q}_{1})\right)_{nm}\label{eq:Over01}
\end{equation}
with
\begin{equation}
L_{nm}(\mathbf{q}_{0})=L_{n_{x}m_{x}}(q_{0x})L_{n_{y}m_{y}}(q_{0y})L_{n_{z}m_{z}}(q_{0z}).\label{eq:L3D}
\end{equation}
It is obvious that the matrices $L(\mathbf{q}_{i})$ have to be identified
with the $L_{i}$ introduced in Eq (\ref{eq:LUR}). The matrix $L_{nm}(\mathbf{q}_{0})$
can be also arranged as a triangular matrix if we take the standard
ordering $n=(n_{x},n_{y},n_{z})$ with $n_{x}=0,\ldots,N_{x}$, $n_{y}=0,\ldots,N_{y}(n_{x})$
and $n_{z}=0,\ldots,N_{z}(n_{x},n_{y})$. With this ordering the inverse
matrix $L_{nm}^{-1}(\mathbf{q}_{0})$ can also be written in analytical
form in terms of the inverse of the 1D quantities given in Appendix
\ref{sec:AppendixB}
\begin{equation}
L_{nm}^{-1}(\mathbf{q}_{0})=L_{n_{x}m_{x}}^{-1}(q_{0x})L_{n_{y}m_{y}}^{-1}(q_{0y})L_{n_{z}m_{z}}^{-1}(q_{0z}).\label{eq:LI3D}
\end{equation}
 The determinant of $\mathcal{R}$ is given by the product of the
determinant of two lower triangular matrices
\[
\det\mathcal{R}=\left(\det\left(L(\mathbf{q}_{0})\right)\right)^{*}\det\left(L(\mathbf{q}_{1})\right).
\]
Given the lower triangular structure of the $L(\mathbf{q}_{i})$ matrices,
their determinant is just the product of the elements in the diagonal
$\det\left(L(\mathbf{q}_{i})\right)=\prod'_{n_{x,}n_{y},n_{z}}q_{ix}^{n_{x}+1/2}q_{iy}^{n_{y}+1/2}q_{iz}^{n_{z}+1/2}$
where the product is restricted to those values of $n_{x},$$n_{y}$
and $n_{z}$ compatible with the definition of the basis (typically,
some energy condition, $\sum_{n_{x},n_{y},n_{z}}\hbar\omega_{x}(n_{x}+1/2)+\hbar\omega_{y}(n_{y}+1/2)+\hbar\omega_{z}(n_{z}+1/2)<E_{0}$.
The explicit form of the LU decomposition of the 1D $\mathcal{R}$
immediately suggests the introduction of the bi-orthogonal states
\begin{equation}
\langle x|m)_{1}=e^{-\frac{x^{2}}{2B^{2}}}\bar{\varphi}_{m}(x,B)=e^{-\frac{x^{2}}{2B^{2}}}\sum_{m'}L_{mm'}^{-1}(q_{1})\bar{\varphi}_{m'}(x,b_{1})\label{eq:BI1}
\end{equation}
and 
\begin{equation}
_{0}(n|x\rangle=e^{-\frac{x^{2}}{2B^{2}}}\bar{\varphi}_{n}^{*}(x,B)=e^{-\frac{x^{2}}{2B^{2}}}\sum_{n'}L_{nn'}^{*-1}(q_{0})\bar{\varphi}_{n'}^{*}(x,b_{0})\label{eq:BI0}
\end{equation}
such that $_{0}(n|m)_{1}=\delta_{nm}$. Clearly, bras and kets of
the bi-orthogonal states turn out to be connected by hermitian conjugation
and therefore they form an unique set of orthogonal states thanks
to the special properties of the HO states. This unique set is just
a set of HO wave functions with oscillator length $B$. 

The matrix elements of one-body momentum-independent operators $\hat{O}$
can be expressed in terms of the matrix elements computed with the
orthogonal basis with oscillator length $\boldsymbol{B}=(B_{x},B_{y},B_{z})$
\begin{equation}
_{0}\langle n|\hat{O}|m\rangle_{1}=\left(L^{*}(\mathbf{q}_{0})O_{\boldsymbol{B}}L^{T}(\mathbf{q}_{1})\right)_{nm}\label{eq:onebME}
\end{equation}
Here $O_{\boldsymbol{B}}$ is the matrix of the matrix elements of
$\hat{O}$ computed with the HO basis with lengths $\boldsymbol{B}$.
In the case of two-body momentum independent operators like the central
or Coulomb potentials the generalization is again straightforward
\begin{align}
_{0}\langle nm|\hat{\upsilon}|pq\rangle_{1} & =\sum_{rstu}L^{*}(\mathbf{q}_{0})_{nr}L^{*}(\mathbf{q}_{0})_{ms}\langle rs|\hat{\upsilon}|tu\rangle_{\boldsymbol{B}}\label{eq:twobME}\\
 & L(\mathbf{q}_{1})_{pt}L(\mathbf{q}_{1})_{qu}.
\end{align}
where $\langle rs|\hat{\upsilon}|tu\rangle_{\boldsymbol{B}}$ are
the matrix elements of the two-body potential computed with HO states
with length $\boldsymbol{B}$.

For the evaluation of momentum dependent operators like the kinetic
energy or the spin-orbit potential the easiest way is to use recursion
relations like
\begin{equation}
\frac{\partial}{\partial x}\widetilde{\varphi}_{n}(x)=\frac{1}{\sqrt{2}B}\left(q_{0}^{2}\sqrt{n}\widetilde{\varphi}_{n-1}(x)-q_{1}^{2}\sqrt{n+1}\widetilde{\varphi}_{n+1}(x)\right)\label{eq:recrel1}
\end{equation}
 and
\begin{align}
\frac{\partial^{2}}{\partial x^{2}}\widetilde{\varphi}_{n}(x) & =\frac{1}{2B^{2}}(q_{0}^{4}\sqrt{n(n-1)}\widetilde{\varphi}_{n-2}(x)\label{eq:recr2}\\
 & -q_{0}^{2}q_{1}^{2}(2n+1)\widetilde{\varphi}_{n}(x)\nonumber \\
 & +q_{1}^{4}\sqrt{(n+1)(n+2)}\widetilde{\varphi}_{n+2}(x))\nonumber 
\end{align}
where$\widetilde{\varphi}_{n}(x)=e^{-\frac{x^{2}}{2b_{1}^{2}}}\bar{\varphi}_{n}(x,B)$.
For instance, the matrix elements of the one-body kinetic energy operator
are given by 
\[
_{0}\langle n|\hat{T}|m\rangle_{1}=\left(L^{*}(\mathbf{q}_{0})T_{\boldsymbol{B}}L^{T}(\mathbf{q}_{1})\right)_{nm}
\]
where the matrix elements $T_{\boldsymbol{B}}$ are computed in the
traditional way, using HO with oscillator lengths $\boldsymbol{B}$
but using the recursion relations of Eqs (\ref{eq:recrel1}) and (\ref{eq:recr2})
instead of the traditional ones. The same recursion relations can
be used in the evaluation of the matrix elements of the two-body spin-orbit
potential. Let us also mention that the corresponding formulas for
the two dimensional HO wave functions often used in axially symmetric
codes can be found in Appendix \ref{sec:AppendixC}.

In the application of the present formalism to the case where density
dependent interactions like Skyrme or Gogny are used we have to use
a prescription for the density dependent term \citep{BONCHE1990466,RoG02d,Rob07,Rob10a}.
The prescription used is the so called overlap prescription that amounts
to use the density
\[
\rho_{\text{ov}}(\vec{r})=\frac{\langle\phi_{0}|\hat{\rho}(\vec{r})|\phi_{1}\rangle}{\langle\phi_{0}|\phi_{1}\rangle}=\sum_{k,l=1}^{N}\varphi_{k}^{*}(\vec{r};b_{0})\varphi_{l}(\vec{r};b_{1})\rho_{lk}^{01}
\]
in the density dependent term of the interaction. 

As mentioned in the previous section (Eqs (\ref{eq:U0tilde}) and
(\ref{eq:U1tilde})) it is convenient to introduce the matrices 
\begin{align}
\tilde{V}_{i} & =L^{+}(\mathbf{q}_{i})\bar{V}_{i}L(\mathbf{q}_{i})\label{eq:TildeV}\\
\tilde{U}_{i} & =\left(L^{*}(\mathbf{q}_{i})\right)^{-1}\bar{U}_{i}L(\mathbf{q}_{i})\label{eq:TildeU}
\end{align}
and
\begin{equation}
\tilde{A}=\tilde{U}_{0}^{T}\tilde{U}_{1}^{*}+\tilde{V}_{0}^{T}\tilde{V}_{1}^{*}=L^{T}(\mathbf{q}_{0})AL^{*}(\mathbf{q}_{1})\label{eq:TildeA}
\end{equation}
They allow to simplify the expression of the overlap to
\begin{equation}
\langle\phi_{0}|\phi_{1}\rangle=\sqrt{\det\tilde{A}}.\label{eq:OverlapTilde}
\end{equation}
This formula is not only simpler than the original one but allows
to avoid a common problem in typical applications: the exceedingly
large or small values of $\det L(\mathbf{q})$ can overflow or underflow
the floating point computer representation of real numbers. We will
discuss this problem in the following Section. Using the definitions
of Eqs (\ref{eq:TildeV}) and (\ref{eq:TildeU}) can also introduce
the density matrix contraction of Eq (\ref{eq:rho10F-1}) to express
the overlap of one body operators in Eq (\ref{eq:OneBodyOverlap})
as 
\begin{equation}
\frac{\langle\phi_{0}|\hat{O}|\phi_{1}\rangle}{\langle\phi_{0}|\phi_{1}\rangle}=\sum_{k,l=1}^{N}O_{kl}^{01}\rho_{lk}^{01}=\textrm{Tr}[O_{\boldsymbol{B}}\tilde{\rho}^{01}]\label{eq:Ov1b}
\end{equation}
given in terms of the matrix elements of the operator in the HO basis
with oscillator lengths $\boldsymbol{B}=(B_{x},B_{y},B_{z})$, i.e.
there is no need to consider additional formulas for matrix element
overlap. The same applies to the calculation of two-body terms like
the overlap of the potential energy Eq (\ref{eq:Twobodyoverlap})
so that one can write finally 
\begin{equation}
\frac{\langle\phi_{0}|\hat{\upsilon}|\phi_{1}\rangle}{\langle\phi_{0}|\phi_{1}\rangle}=\frac{1}{2}\mathrm{Tr}[\tilde{\Gamma}^{01}\tilde{\rho}^{01}]-\frac{1}{2}\mathrm{Tr}[\tilde{\Delta}^{01}\tilde{\bar{\kappa}}^{01}]\label{eq:Ov2b}
\end{equation}
with the HF potential 
\begin{equation}
\tilde{\Gamma}_{ik}^{01}=\frac{1}{2}\sum\tilde{\upsilon}_{ijkl}^{\boldsymbol{B}}\tilde{\rho}_{lj}^{01}\label{eq:HFGamma}
\end{equation}
 and pairing field 
\begin{equation}
\tilde{\Delta}_{ij}^{01}=\frac{1}{2}\sum\tilde{\upsilon}_{ijkl}^{\boldsymbol{B}}\tilde{\kappa}_{kl}^{01}\label{eq:Delta01}
\end{equation}
 computed with matrix elements of the potential in the HO basis with
lengths $\boldsymbol{B}$(except for the momentum dependent terms
of the potential where, additionally, one has to use the modified
recursion relations of Eqs (\ref{eq:recrel1}) and (\ref{eq:recr2}).
For the spatial overlap density finally obtain
\[
\rho_{\text{ov}}(\vec{r})=\sum_{k,l=1}^{N}\varphi_{k}^{*}(\vec{r};B)\varphi_{l}(\vec{r};B)\tilde{\rho}_{lk}^{01}
\]

To summarize this section, the calculation of overlaps of operators
between HFB states expressed in HO basis with different oscillator
lengths involves 
\begin{enumerate}
\item Computing matrix elements of the operators with a set of HO functions
with oscillator length parameters $\boldsymbol{B}$ given by Eq (\ref{eq:Bosc}).
\item Computing the $\tilde{U}_{i}$ and $\tilde{V}_{i}$ Bogoliubov amplitudes
Eqs (\ref{eq:TildeV}) and (\ref{eq:TildeU}) and $\tilde{A}$ in
Eq (\ref{eq:TildeA}).
\item Computing the density matrix and pairing tensor contractions of Eqs
(\ref{eq:rho10F-1}), (\ref{eq:kappa01F-1}) and (\ref{eq:kappa10F-1}).
\item Evaluating the HF potential Eq (\ref{eq:HFGamma}) and pairing field
(\ref{eq:Delta01}).
\item Evaluate overlaps for one-body Eq (\ref{eq:Ov1b}) and two-body Eq
(\ref{eq:Ov2b}) operators.
\end{enumerate}
Please note that Eqs (\ref{eq:TildeA}), (\ref{eq:rho10F-1}), (\ref{eq:kappa01F-1})
and (\ref{eq:kappa10F-1}) have the traditional form but in terms
of $\tilde{U}_{i}$ and $\tilde{V}_{i}$ Bogoliubov amplitudes Eqs
(\ref{eq:TildeV}) and (\ref{eq:TildeU}). Therefore, the modifications
required to implement the formalism described in an existing computer
code are minimal and easy to implement.

\section{Application of the method}

\begin{figure}
\includegraphics[width=0.95\columnwidth]{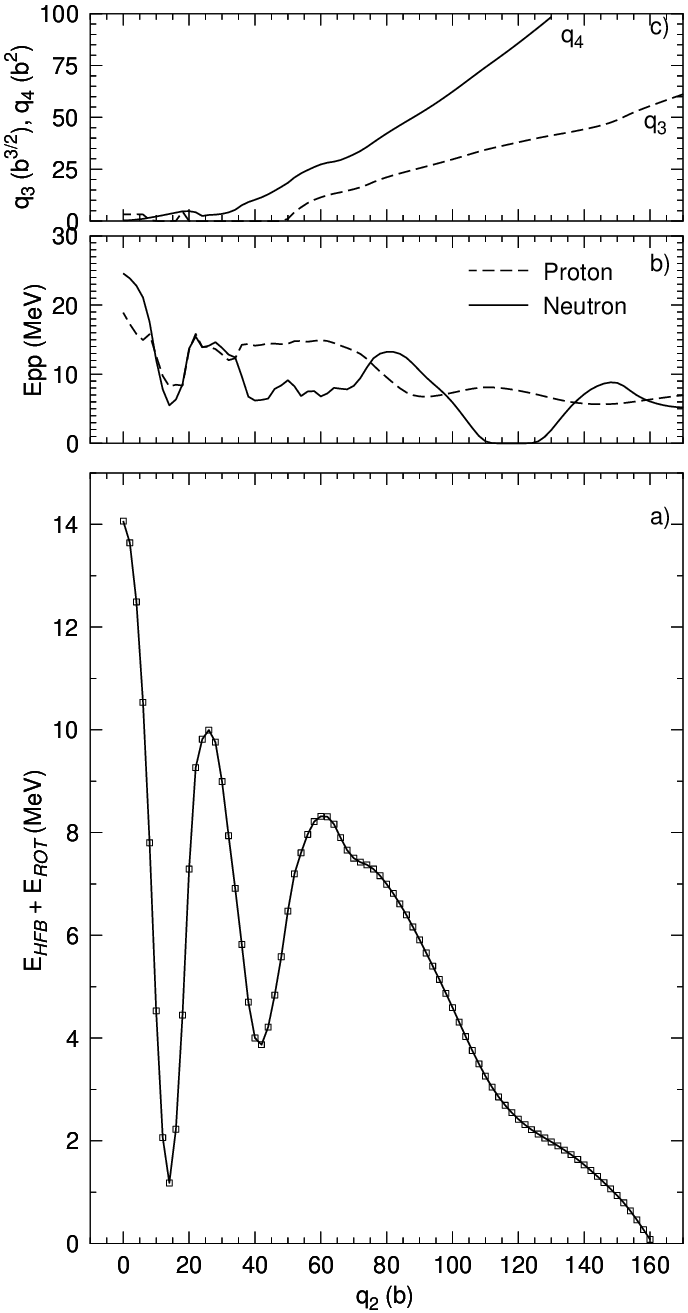}\caption{In the lower panel, the potential energy surface (including the rotational
energy correction) of $^{238}$Pu is depicted as a function of the
quadrupole moment $q_{20}$ (in barns). The energy has been shifted
by 1800 MeV. In the middle panel the particle-particle (pairing) correlation
energy for both protons and neutrons. In the upper panel, the octupole
and hexadecapole moments of each HFB configuration as a function of
$q_{20}$. \label{fig:PES-of-Pu}}
\end{figure}

In this section the formalism will be used to compute the overlaps
between the members of the set of wave functions $|\phi(q_{20})\rangle$
entering the fission path of the nucleus $^{238}$Pu. Traditionally,
those wave functions are computed in a HO basis with oscillator lengths
tailored to the deformation $q_{20}$ and obtained by minimizing the
HFB energy as a function of the oscillator lengths. The results will
be compared to the ones obtained by a blind application of the standard
formulas. To motivate the discussion, we display in Fig \ref{fig:PES-of-Pu}
the potential energy surface (PES) obtained as a function of the axial
quadrupole moment $q_{20}$ with the Gogny force D1M{*} \citep{D1M_S,Boquera21}.
In the calculation we use an axial HO basis with 18 shells in the
perpendicular direction and 27 shells in the $z$ direction. 

In Fig \ref{fig:PES-of-Pu}, apart from the HFB energy plus the rotational
correction {[}panel a){]} other relevant quantities like the particle-particle
pairing energy for protons and neutrons {[}panel b){]}, the octupole
and hexadecapole moments {[}panel c){]} are shown. We are dealing
with a standard actinide with a deformed ground state, a fission isomer
at $q_{20}=44$ b and two barriers (inner and outer). At $q_{20}=50$
b, the HFB solution starts breaking reflection symmetry and a non-zero
octupole moment develops. 
\begin{figure}

\includegraphics[width=0.95\columnwidth]{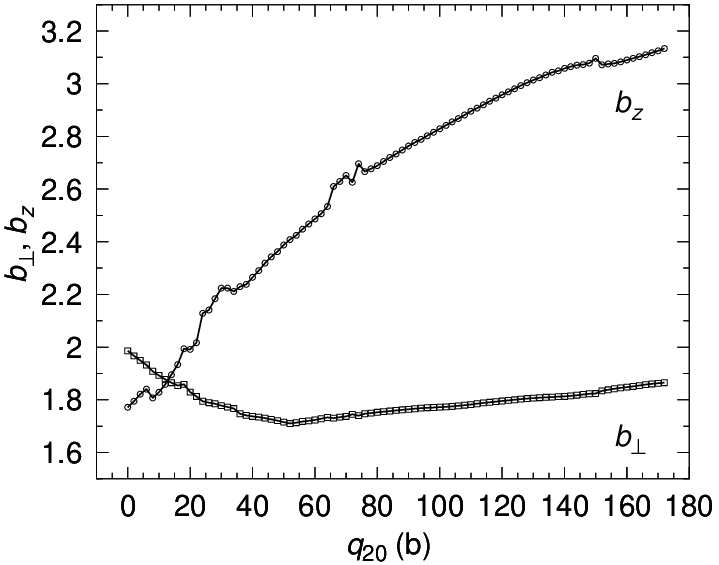}\caption{Oscillator lengths $b_{\perp}$ and $b_{z}$ for each HFB configuration
$|\phi(q_{20})\rangle$. \label{fig:Oscillator-lengths}}

\end{figure}

The oscillator lengths $b_{\perp}$ and $b_{z}$ obtained by minimizing
the HFB energy for each value of the quadrupole moment $q_{20}$ are
depicted in Fig \ref{fig:Oscillator-lengths}. The abrupt changes
observed at some $q_{20}$ values are due to coexisting minima in
the HFB energy $E_{HFB}(q_{20})$ as a function of $b_{\perp}$ and
$b_{z}$. We could have chosen the oscillator lengths as to obtain
a smoother curve in Fig \ref{fig:Oscillator-lengths} but we have
preferred to leave that way as this is the typical outcome of an automatized
procedure. As shown in Fig \ref{fig:Oscillator-lengths}, $b_{z}$
increases linearly with $q_{20}$ whereas $b_{\perp}$ remains roughly
constant in the whole interval. It is also evident that there are
large variations in $b_{z}$ depending on the value of the quadrupole
moment. 

To illustrate the results obtained with the present formalism, we
have chosen three emblematic values of $q_{20}$, namely 40 b, 80
b and 120 b and varied the corresponding optimal oscillator lengths
by $\pm$0.05 fm to obtain different HFB solutions. Those solutions
have roughly the same energy and other observables as the starting
one, indicating that the HFB wave function is essentially the same
in all the cases. However, as the oscillator length parameters have
slightly changed, the HFB amplitudes must be slightly different as
to absorb the changes in the basis parameters. As a consequence, we
expect the traditional formula for the overlap (i.e. the one not taking
into account that the bases are different) to give values differing
from one. On the other hand, the formalism introduced here should
provide an overlap close to one. The results obtained are summarized
in Table \ref{tab:The-optimal-oscillator}. They clearly show how
the traditional overlap formula is wrong whereas the present formalism
provide the expected results, namely overlaps very close to one. 

\begin{table*}
\begin{tabular}{|c|c|c|c|c|c|c|c|c|c|}
\hline 
Config & \multicolumn{3}{c|}{$q_{20}=40$b} & \multicolumn{3}{c|}{$q_{20}=80$b} & \multicolumn{3}{c|}{$q_{20}=120$b}\tabularnewline
\hline 
\hline 
 & $E_{HFB}$ & Eq (\ref{eq:OverlapTilde}) & Trad & $E_{HFB}$ & Eq (\ref{eq:OverlapTilde}) & Trad & $E_{HFB}$ & Eq (\ref{eq:OverlapTilde})  & Trad\tabularnewline
\hline 
Optimal & -1792.104 & 1.0 & 1.0 & -1788.366 & 1.0 & 1.0 & -1792.883 & 1.0 & 1.0\tabularnewline
\hline 
$b_{z}+0.05$ & -1792.040 & 0.999 & 0.721 & -1788.317 & 0.999 & 0.747 & -1792.841 & 0.999 & 0.766\tabularnewline
\hline 
$b_{z}+0.10$ & -1791.931 & 0.998 & 0.279 & -1788.216 & 0.997 & 0.322 & -1792.746 & 0.998 & 0.354\tabularnewline
\hline 
$b_{z}-0.05$ & -1792.064 & 0.998 & 0.709 & -1788.324 & 0.999 & 0.735 & -1792.841 & 0.999 & 0.755\tabularnewline
\hline 
$b_{z}-0.10$ & -1791.838 & 0.993 & 0.243 & -1788.155 & 0.994 & 0.283 & -1792.841 & 0.996 & 0.315\tabularnewline
\hline 
$b_{\perp}+0.05$ & -1792.099 & 0.999 & 0.820 & -1788.360 & 0.999 & 0.830 & -1792.873 & 0.999 & 0.803\tabularnewline
\hline 
$b_{\perp}-0.05$ & -1792.099 & 0.999 & 0.813 & -1788.366 & 0.999 & 0.824 & -1792.868 & 0.999 & 0.797\tabularnewline
\hline 
\end{tabular}\caption{Overlaps between different HFB solutions obtained with slightly different
oscillator lengths. The optimal oscillator lengths ($b_{\perp},b_{z})$
are (1.75, 2.25), (1.75, 2.75) and (1.80,2.95) for $q_{20}=40$ b,
80 b and 120 b, respectively. The column denoted as ``Trad'' corresponds
to the traditional calculation of the overlaps without taking into
account the effect of the bases (Onishi formula). \label{tab:The-optimal-oscillator}}
\end{table*}

\begin{figure}

\includegraphics[width=0.95\columnwidth]{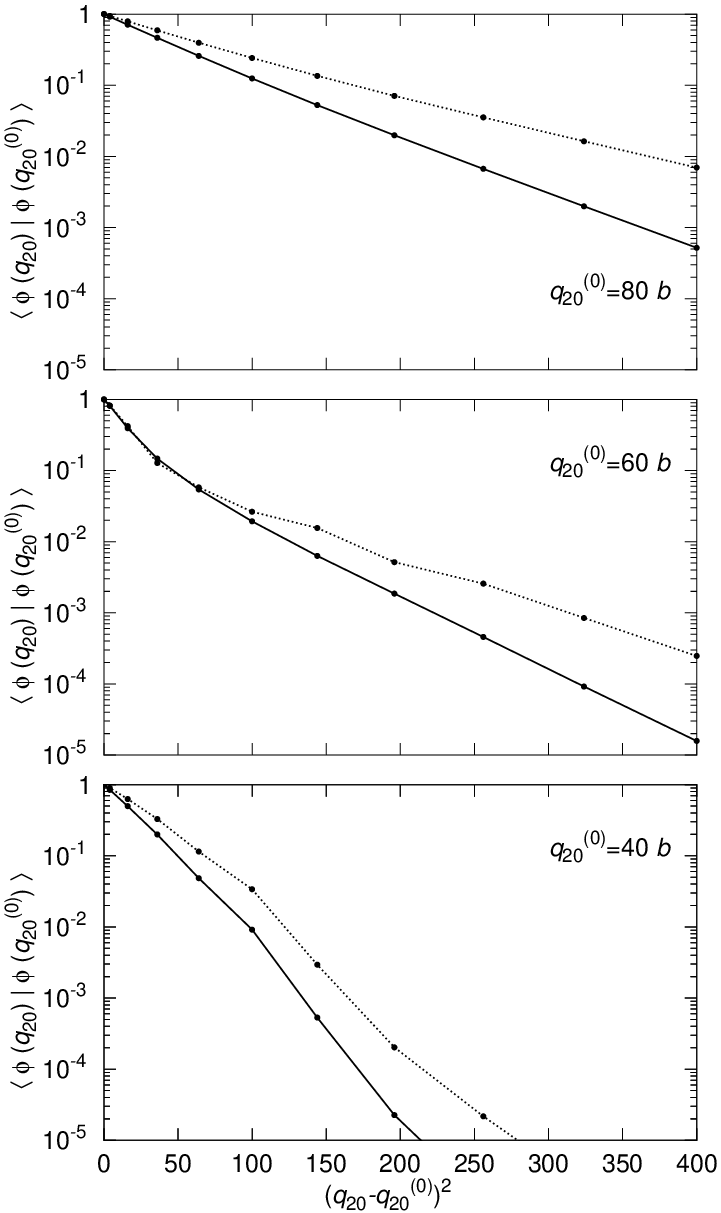}\caption{Overlaps $\langle\phi(q_{20})|\phi(q_{20}^{(0)})\rangle$ as a function
of $\left(q_{20}-q_{20}^{(0)}\right)^{2}$ for different representative
values of $q_{20}^{(0)}$. \label{fig:Overlaps}}
\end{figure}

To finalize this section, let us consider now the overlaps $\langle\phi(q_{20})|\phi(q_{20}^{(0)})\rangle$
with $q_{20}^{(0)}$= 40 b, 60 b and 80 b as a function of $q_{20}$.
They are plotted in logarithmic scale in Fig \ref{fig:Overlaps} as
a function of $\left(q_{20}-q_{20}^{(0)}\right)^{2}$, motivated by
the fact that in the Gaussian overlap approximation $\langle\phi(q_{20})|\phi(q_{20}^{(0)})\rangle\approx\exp\left[-\gamma(q_{20})\left(q_{20}-q_{20}^{(0)}\right)^{2}\right]$.
The exact overlaps computed with Eq (\ref{eq:OverlapTilde}) are represented
with a solid line whereas the wrong ones, computed with the Onishi
formula, are plotted using dotted lines. There are several salient
features worth to mention: first, the overlaps given by the Onishi
formula are almost always larger than the correct ones. The reason
is that as the oscillator lengths are adapted to the quadrupole moment,
the basis is partially responsible for generating quadrupole deformation
and therefore the Bogoliubov amplitudes of neighboring configurations
change less than if the oscillator lengths of the basis were kept
constant. The feature shows some exceptions in the limited $q_{20}$
region in the $q_{20}^{(0)}=60$ b case where the oscillator lengths
as a function of $q_{20}$ show a less smooth behavior as in the other
cases. Second, the width of the Gaussian is larger with Eq (\ref{eq:OverlapTilde})
than with the Onishi formula except in the $q_{20}^{(0)}=60$ b case.
Third, in the $q_{20}^{(0)}=60$ b case Eq (\ref{eq:OverlapTilde})
provides far more smooth results than the Onishi formula indicating
that Eq (\ref{eq:OverlapTilde}) is able to absorb the changes is
the oscillator lengths taking place in this case. Fourth, the Gaussian
overlap formula seems to be rather inaccurate in the $q_{20}^{(0)}=40$
b and 60 b cases as the behavior of the overlap in the given scales
depart from the expected straight line. Finally, let us briefly discuss
the large and small values of $\det L_{i}$ found in the calculation
of the overlaps. As a typical example, consider $q_{20}^{(0)}=40$
b and $q_{20}=60$ b. In this case, $\det L_{0}=1.8\times10^{186}$
and $\det L_{1}=6.8\times10^{-208}$ to give $\det\mathcal{R}=1.3\times10^{-21}.$
At $q_{20}$=64 b $\det\mathcal{R}=1.2\times10^{-30}$ but the overlap
is still relatively relevant with a value of $2.03\times10^{-4}.$
However, at $q_{20}$=66 b $\det L_{1}$ underflows the 64 bit floating
point representation used in the calculations and $\det\mathcal{R}=0$
to machine accuracy. A bit further, at $q_{20}=68$ b the determinant
$\det L_{1}$ is the one who overflows and the computer value for
$\det\mathcal{R}$ is undefined. However, using Eqs (\ref{eq:Atilde})
and  (\ref{eq:Overtilde}) the evaluation of the overlap proceeds
smoothly and, what is more important, unattended. 

\begin{acknowledgments}
This work has been supported by the Spanish Ministerio de Ciencia,
Innovación y Universidades and the European regional development fund
(FEDER), grants No PGC2018-094583-B-I00.
\end{acknowledgments}

\appendix

\section{The dilatation operator\label{sec:The-dilatation-operator}}

The dilatation operator in 1D,$\hat{\mathcal{T}}(a)$ satisfies $\langle x|\hat{\mathcal{T}}(a)=\langle ax|$
and therefore $\langle x|\hat{\mathcal{T}}(a)|\Psi\rangle=\Psi(ax).$
In one dimension, the operator is given by 
\[
\hat{\mathcal{T}'}(a)=e^{\tau x\partial_{x}}
\]
with $a=e^{\tau}.$ As it stands, the operator is not unitary as the
exponent is not an anti-hermitian operator. As it can be easily checked,
just adding $\tau+1/2$ to the exponent do the trick, and the unitary
dilation operator is given by 
\[
\hat{\mathcal{T}}(a)=e^{\tau\left(x\partial_{x}+\frac{1}{2}\right)}
\]
Now $\hat{\mathcal{T}}(a)f(x)=e^{\tau/2}f(e^{\tau}x)$. The extra
factor $\exp(\tau/2)$ is usually absorbed by the normalization constant,
as it is the case for the harmonic oscillator wave function 

\section{Transformation coefficient $L$ in one dimension\label{sec:AppendixB}}

In this Appendix we establish the explicit expression of the $L$
coefficients in the expansion of the restricted 1D HO wave functions
\[
\bar{\varphi}_{n}(x,b_{0})=\sum_{m=0}^{n}L_{nm}(q_{01})\bar{\varphi}_{m}(x,b_{1})
\]
with $q_{01}=\frac{b_{1}}{b_{0}}$ . They are given by
\begin{equation}
L_{nm}(q)=\Delta_{n,m}\frac{\left(\frac{n!}{m!}\right)^{1/2}}{2^{(n-m)/2}\left(\frac{n-m}{2}\right)!}(q^{2}-1)^{\frac{n-m}{2}}q^{m+1/2}\label{eq:Lnm}
\end{equation}
where $\Delta_{n,m}=\frac{1}{2}[1+(-)^{n+m}]$ is the ``parity''
Kronecker symbol. The $L_{nm}$ also satisfy $L_{nm}=0$ if $m>n$
due to the factorial in the denominator and therefore they are the
matrix elements of a lower triangular matrix. The expression of Eq.\textasciitilde\ref{eq:Lnm}
is easily obtained by using the generating function of the Hermite
polynomials
\[
\sum_{n}\frac{t^{n}}{n!}H_{n}(x)=e^{2xt-t^{2}}.
\]
The inverse of the 1D matrix $L$ can be found by inspection
\begin{equation}
L_{mr}^{-1}(q)=\Delta_{m,r}\frac{(-)^{\frac{m-r}{2}}\left(\frac{m!}{r!}\right)^{1/2}}{2^{(m-r)/2}\left(\frac{m-r}{2}\right)!}(q^{2}-1)^{\frac{m-r}{2}}q^{-(m+1/2)}.\label{eq:ILmr}
\end{equation}
Due to the triangular form, the determinant of $L$ can also be computed
right away: assuming the dimension of the matrix is $N$ (i.e. $m=0,\ldots,N-1$)
\[
\det L=\prod_{m=0}^{N-1}q^{m+1/2}=q^{N^{2}/2}
\]

\section{Transformation coefficient L in two dimensions\label{sec:AppendixC}}

The generalization of the results of Appendix \ref{sec:AppendixB}
to the two dimensional case that one encounters in harmonic oscillator
basis with axial symmetry is straightforward. The 2D HO wave function
is defined in terms of the $n_{\perp}$ and $m$ quantum numbers as
\[
\phi_{n_{\perp}m}(\vec{r},b)=e^{-\frac{1}{2}r_{\perp}^{2}}\bar{\phi}_{n_{\perp}m}(\vec{r},b)
\]
with
\[
\bar{\phi}_{n_{\perp}m}(\vec{r},b)=\mathcal{N}_{\perp}\left(\frac{r_{\perp}}{b}\right)^{|m|}L_{n_{\perp}}^{|m|}\left(\frac{r_{\perp}^{2}}{b^{2}}\right)e^{im\varphi}
\]
and $\mathcal{N}_{\perp}=\frac{1}{b\sqrt{\pi}}\left(\frac{n_{\perp}!}{(n_{\perp}+|m|)!}\right)^{1/2}$.
The generating function is in this case
\[
\exp\left(\frac{2\vec{r}\vec{t}}{b}-t^{2}\right)=\sum_{n_{\perp}m}\kappa_{n_{\perp}m}^{*}(\vec{t})\bar{\phi}_{n_{\perp}m}(\vec{r},b)
\]
where $\vec{t}$ is a two-dimensional vector and the coefficients
in the linear combination are given by
\[
\kappa_{n_{\perp}m}^{*}(\vec{t})=(-)^{n_{\perp}}\frac{\sqrt{\pi}b}{\left(n_{\perp}!(n_{\perp}+|m|)!\right)^{1/2}}t_{\perp}^{2n_{\perp}+|m|}e^{-im\varphi_{t}}
\]
Expanding the identity 
\[
\exp\left(\frac{2\vec{r}\vec{t}}{b}-t^{2}\right)=\exp\left(\frac{2\vec{r}(q\vec{t})}{b'}-q^{2}t^{2}\right)\exp\left(\left(q^{2}-1\right)t^{2}\right)
\]
in powers of $t_{\perp}$and $\varphi_{t}$ and equating equal powers
in both sides we obtain
\[
\bar{\phi}_{n_{\perp}m}(\vec{r},b)=\sum_{n'_{\perp}}L_{n_{\perp}n'_{\perp}}^{|m|}(q)\bar{\phi}_{n'_{\perp}m}(\vec{r},b')
\]
with 
\[
L_{n_{\perp}n'_{\perp}}^{|m|}(q)=q^{2n'_{\perp}+|m|+1}\frac{(1-q^{2})^{n_{\perp}-n'_{\perp}}}{(n_{\perp}-n'_{\perp})!}\left(\frac{n_{\perp}!(n_{\perp}+|m|)!}{n'_{\perp}!(n'_{\perp}+|m|)!}\right)^{1/2}
\]
and $q=b'/b.$ It is usually more convenient to express the quantum
numbers in terms of $N=2n_{\perp}+|m|$ and $m$. Using them we obtain
\begin{align*}
L_{(N,m)(N',m')}(q) & =\frac{\delta_{mm'}(-)^{\frac{N-N'}{2}}}{\left(\frac{(N-N')}{2}\right)!}\left(\frac{\left(\frac{N\pm m}{2}\right)!}{\left(\frac{N'\pm m'}{2}\right)!}\right)^{1/2}\\
 & \times(1-q^{2})^{\frac{N-N'}{2}}q^{(N'+1)}
\end{align*}
and for the inverse 
\begin{align*}
L_{(N,m)(N',m')}^{-1}(q) & =\frac{\delta_{mm'}(-)^{\frac{N-N'}{2}}}{\left(\frac{(N-N')}{2}\right)!}\left(\frac{\left(\frac{N\pm m}{2}\right)!}{\left(\frac{N'\pm m'}{2}\right)!}\right)^{1/2}\\
 & \times(1-q^{2})^{\frac{N-N'}{2}}q^{-(N+1)}
\end{align*}
as can be easily obtained by inspection. In the above formulas $\left(\frac{N\pm m}{2}\right)!=\left(\frac{N-m}{2}\right)!\left(\frac{N+m}{2}\right)!$

\bibliographystyle{apsrev4-2}
%

\end{document}